\documentclass{emulateapj}
\defcitealias{seok13}{SKO13}
\defcitealias{pin11}{PG11}

\shorttitle{}
\shortauthors{Seok et al.}

\begin{document}

\title{Dust cooling in supernova remnants in the Large Magellanic Cloud}

\author{Ji Yeon Seok\altaffilmark{1,2}, Bon-Chul Koo\altaffilmark{3}, and
Hiroyuki Hirashita\altaffilmark{2}}

\altaffiltext{1}{Department of Physics and Astronomy, University of Missouri, Columbia, MO 65211, USA, seokji@missouri.edu}
\altaffiltext{2}{Institute for Astronomy and Astrophysics, Academia Sinica, 
P.O. Box 23-141, Taipei 10617, Taiwan}
\altaffiltext{3}{Department of Physics and Astronomy, Seoul National University, 
Seoul 151-742, Korea}

\begin{abstract}

The infrared-to-X-ray (IRX) flux ratio traces the relative importance of dust cooling
to gas cooling in astrophysical plasma such as supernova remnants (SNRs). 
We derive IRX ratios of SNRs in
the LMC using $Spitzer$ and $Chandra$ SNR survey data and
compare them with those of Galactic SNRs.
IRX ratios of all the SNRs in the sample are found to be moderately greater than unity,
indicating that dust grains are a more efficient coolant
than gas although gas cooling may not be negligible. 
The IRX ratios of the LMC SNRs are
systematically lower than those of the Galactic SNRs. As both dust cooling and
gas cooling pertain to the properties of the interstellar medium, the lower IRX ratios
of the LMC SNRs may reflect the characteristics of the LMC, and the lower dust-to-gas
ratio (a quater of the Galactic value) is likely to be the most significant factor. The observed
IRX ratios are compared with theoretical predictions that yield IRX ratios
an order of magnitude larger.
This discrepancy may originate from the dearth
of dust in the remnants due to either the local variation of the dust abundance in the preshock
medium with respect to the canonical abundance or the dust destruction in the postshock
medium. The non-equilibrium ionization cooling of hot gas, in particular for young SNRs, may also
cause the discrepancy.
Finally, we discuss implications for the dominant cooling mechanism of SNRs in
low-metallicity galaxies. 

\end{abstract}


\section{Introduction}\label{sec:intro}

A dusty medium is known to be cooled either by an infrared (IR)
cooling mechanism via gas--grain collisions referred to as dust cooling or by a
cooling mechanism via atomic processes referred to as gas cooling. In hot
plasma ($\ga10^6$ K), the cooling process is expected to be dominated by dust cooling
\citep[e.g.,][]{ostriker73}. Since the canonical temperature of the swept-up
gas in a supernova remnant (SNR) exceeds $10^6$ K \citep[e.g.,][]{hughes98},
SNRs provide an ideal laboratory to examine dust cooling in the
interstellar medium (ISM).

A measure of the relative strength, the IR-to-X-ray (IRX) flux ratio, is often used as an
indicator of the dominant cooling mechanism in dusty plasma \citep{dwek87,dwek87a}.
Using IR data obtained by the {\it Infrared Astronomical Satellite} ($IRAS$),
IRX ratios of nine Galactic SNRs and four SNRs in the LMC
have been measured by \citet{dwek87a} and \citet{graham87}, respectively, and
both studies found that IRX ratios of the SNRs are significantly greater
than unity, supporting dust grains being the dominant coolant.

As dust and gas cooling pertains to the properties of the ISM such as
dust-to-gas ratio (DGR) or metallicity, IRX ratios in different galaxies or in different environments 
may reveal a distinct tendency. If an SNR or its progenitor has encountered
inhomogeneities in the ISM or local variations of dust abundance in the ambient
medium, e.g., SNRs interacting with nearby molecular
clouds \citep{dwek87a}, IRX ratios may deviate significantly from
the theoretical prediction based on global values.
Also, the variation of IRX ratios could be a probe of the ongoing destruction of dust
\citep[e.g.,][]{dwek08}.

The LMC is the nearest
gas-rich galaxy to our Galaxy, and its metallicity is a half of the Galactic value
\citep[e.g.,][]{pei92}. In this context, a comparison of the IRX ratios of the LMC SNRs
to those of the Galactic SNRs
may confirm environmental effects on the dominant cooling mechanism in the hot
plasma. However, no systematic comparison of IRX ratios between SNRs in the two galaxies
has been made so far, probably due to the insufficient IR and/or X-ray data on the SNRs.  

In the last decade, the {\it Spitzer Space Telescope} and the {\it AKARI} infrared satellite 
have enabled large-scale IR surveys to be performed toward the Galaxy as well as the Magellanic Clouds
\citep[e.g.,][]{meixner06, onaka07}. Using these data, 
statistical studies about the IR emission of SNRs in the Galaxy
\citep[e.g.,][]{lee05, reach06, pin11, jeong12}
and the LMC \citep[e.g.,][]{seok08,seok13} have been carried out.
Moreover, the {\it Chandra X-ray Observatory} provides superb X-ray data on a large
number of SNRs such as the {\it Chandra} SNR
Catalog\footnote{http://hea-www.cfa.harvard.edu/ChandraSNR/}.
The IR and X-ray surveys on SNRs allow us to
carry out a systematic comparison of IRX ratios and to investigate 
environmental effects on this ratio. 

In this paper, we present IRX ratios of LMC SNRs using $Spitzer$ IR fluxes
and compare them with those of Galactic SNRs. Theoretical IRX ratios
for the LMC and the Galaxy are calculated according to dust and gas cooling functions taking
into account relevant DGRs. The IRX ratios of the LMC SNRs show a subtle deviation
from those of the Galactic SNRs, and for some cases, the observed IRX ratios are significantly
lower than the theoretical prediction. We discuss the the possible
origin of the observed trend and its implications.

\section{IRX ratio of a dusty plasma}\label{sec:IRX}

\subsection{Measurement of IRX ratio}\label{subsec:obs}

We define the observational IRX ratio as
\begin{equation}
{\rm IRX~ratio}\equiv F_{\rm IR}/F_{\rm X},
\label{eq:obsirx}
\end{equation}
where $F_{\rm IR}$ is the total IR flux and $F_{\rm X}$ is the $Chandra$ X-ray flux
in 0.3--2.1 keV. We adopt the $Spitzer$ 24 and 70 \micron~fluxes of the LMC SNRs from
\citet[hereafter SKO13]{seok13}
and X-ray fluxes from the $Chandra$ SNR catalog.
Among the 47 SNRs listed in \citetalias{seok13}, 29 SNRs show IR emission at 24 \micron, 
and all but three SNRs are also detected at 70 \micron. 
In addition, X-ray properties (gas temperature and density) of 28 SNRs are listed in \citetalias{seok13},
and 24 out of the 28 SNRs have IR flux measurements.
However, note that we include 19 out of the 24 SNRs in this study for the following reasons: 
IR and X-ray emission of two SNRs (N157B and N158A) are dominated by their pulsar wind nebulae, 
so dust and gas cooling mechanisms are not applicable.
For two other SNRs (DEM L72 and DEM L256),
although their X-ray properties are presented in \citet{klimek10}
their total X-ray fluxes are not given in the paper. Finally, in the case of the SNR in N159,
it is inevitable that its IR and X-ray emissions are contaminated by
a nearby \ion{H}{2} region and LMC X-1.

Among the 19 SNRs,
we estimate $F_{\rm IR}$ of 16 of them by fitting their spectral energy distribution (SED)
with a single-temperature modified blackbody (Equation (1) in \citetalias{seok13}). 
Uncertainties of $F_{\rm IR}$ mostly result from the uncertainties of the photometry,
which is $\sim15\%$--$20\%$ of the total flux.
In the case of the three SNRs (0509--67.5, 0519--69.0, and SN 1987A)
for which 70 \micron~fluxes are not given in \citetalias{seok13},
their $F_{\rm IR}$ are derived on the basis of
the spectrum taken by the Infrared Spectrograph (IRS)
on board $Spitzer$ \citep[see more in Table \ref{tab:flx}]{dwek10, bwill11}.

For all the SNRs but two (DEM L205 and DEM L241),
we adopt X-ray fluxes from the $Chandra$ SNR catalog (see Table \ref{tab:flx}).
Adopting the soft-band (0.3--2.1 keV) fluxes from the $Chandra$ catalog,
we expect to exclude a hard X-ray component mainly dominated by synchrotron emission. 
In fact, the soft-band fluxes are nearly the same as the wide-band (0.3--10.0 keV) fluxes
except for (young) Crab-like SNRs where synchrotron emission dominates their SEDs
(such as N157B and N158A that we likewise exclude above).
Although there are three Crab-like SNRs
(0453--68.5, SNR in N206, and DEM L241) in our sample,
all are mature enough for their SNR shocks to have encountered the ambient medium
and to produce thermal emission contributing most of the X-ray emission.
For a few (young) SNRs such as N103B or SN 1987A,
however, the X-ray spectra indicate multiple
components, possibly including a contribution from SN ejecta,
which need to be distinguished (see section \ref{sec:unc}). 
Uncertainties of the X-ray fluxes mostly arise
when correcting the interstellar absorption in general,
so that the X-ray fluxes of the LMC SNRs should be relatively secure compared
with those of the Galactic SNRs.
We suppose that the X-ray fluxes are uncertain by a factor of a few, at most
\citep[e.g.,][]{seok08, dwek10} and assume 30\% of the X-ray flux as its uncertainty
en bloc.

To compare LMC and Galactic SNRs,
we accumulate IR fluxes of Galactic SNRs in the literature. 
Recently, \citet[hereafter \citetalias{pin11}]{pin11} detected 39 out of 121 SNRs within 
the $Spitzer$ MIPSGAL survey at 24 and 70 \micron~and measured their fluxes. 
For 13 of these 39 SNRs, the total IR fluxes and IRX ratios are also measured.
We consider that the 70 \micron~fluxes of two SNRs
(Kes 79 and G337.2--0.7) are quite uncertain
due to their low brightness in the MIPS 70 \micron,
so they are excluded from our study.
Moreover, we notice that most SNRs with the measurement of IRX ratios in \citetalias{pin11} are known to interact 
with molecular clouds,
which do not usually show a spatial resemblance between IR and X-ray
(see below). To supplement Galactic SNRs for this study,
eight Galactic SNRs including the historical SNRs such
as those of Kepler or Tycho, are additionally included.
In total, there are 19 Galactic SNRs used in this work.
Since \citetalias{pin11} derive IRX ratios using $Chandra$ wide-band fluxes
\citepalias[see Table 6 in][]{pin11}, 
we newly derive their IRX ratios using the soft-band fluxes for consistency of this paper
although the difference is negligible.

Furthermore, as the spatial resemblance between IR and X-ray morphologies
(IRX resemblance, hereafter)
implies a physical association between dust and hot plasma, it is necessary to
scrutinize IRX ratios taking the IRX resemblance into account.
We classify all SNRs in our sample into three categories (definite, partial, and lack
of resemblance)
based on visual inspection of the $Spitzer$ 24 \micron~and $Chandra$
soft-band images and use the classification for further analysis.

In summary, we estimate IRX ratios of the 19 LMC SNRs and 19 Galactic SNRs
as listed in Table \ref{tab:flx}.
The total IR and X-ray fluxes, plasma temperatures, and the IRX resemblances are given together.
Among the LMC SNRs, \citet{graham87} previously measured IRX ratios of N49, N49B, and N63A
using the $IRAS$ data ($\sim12$, 4, and 12, respectively). While our new measurement for N49
(N49: $11.1\pm2.9$) is consistent with the previously derived ratio, the IRX ratios of the other two SNRs
(N49B: $1.51\pm0.39$ and N63A: $2.96\pm0.76$) differ somewhat from them. This deviation is most
likely due to the contribution from the FIR emission (see section \ref{sec:unc}). 

\begin{deluxetable*}{lccccc}
\tablecolumns{6}
\tablewidth{0pt}
\tablecaption{Observed X-ray and IR fluxes of SNRs}
\tablehead{\colhead{SNR}  & \colhead{$F_{\rm IR}$\tablenotemark{a}} &
	\colhead{$F_{\rm X}$\tablenotemark{b}} & \colhead{IRX Ratio} & 
	\colhead{$T_e$\tablenotemark{c}} &
	 \colhead{IR-X-ray Resemblance \tablenotemark{d}} \\
	\colhead{} & \colhead{($10^{-11}$ erg cm$^{-2}$ s$^{-1}$)} &
	\colhead{($10^{-11}$ erg cm$^{-2}$ s$^{-1}$)} & \colhead{} &\colhead{($10^6$ K)} & 
	\colhead{}} 
	\startdata
	\\
	\multicolumn{6}{l}{LMC Sources} \\
	\hline
	0453--68.5 & $3.84 \pm0.69$ & 1.8 &  $2.12\pm0.57$ & ~3.4  & D\\
	DEM L71 & $3.85\pm0.66$ & 2.6  & $1.50\pm0.39$ & ~7.5  & D\\
	N23 & $3.57\pm0.56$ & 2.2 &  $1.60\pm0.41$ & ~6.5 &D\\
	N103B & $9.17\pm1.36$ & 5.6 & $1.63\pm0.41$ & 11.6 &D\\
	0509--67.5 & $0.85\pm0.43$ & 0.59 & $1.43\pm0.78$ & 23.2 & D\\
	0519--69.0 & $2.10\pm1.27$ & 3.57 & $0.58\pm0.38$ & 17.4 & D\\	
	N132D & $26.3\pm3.7~$& 17.0  &$1.55\pm0.38$ & ~9.5 &D\\
	N49B & $12.1\pm2.0~$& 8.0 & $1.51\pm0.39$ & ~4.2 &D\\
	N49 & $79.2\pm13.5$ & 7.2 & $11.1\pm2.90 $ & ~5.1 &P\\
	DEM L205 &  $5.81\pm0.77$ &  0.048 & $121\pm29.2$ &  ~2.9 & L \\
	SNR in N206 & $1.11\pm0.19$ &  0.44 &  $ 2.52\pm0.66$ & ~4.6 & L \\
	DEM L238 & $0.57\pm0.10$ &  0.072 &   $7.97\pm2.11$ &  ~4.1 & L \\
	SN 1987A\tablenotemark{e} & 0.31--1.55 & 0.12--0.58  & 2.4--3.8 (2.5) & ~5.0 &D\\
	N63A & $76.1\pm12.1$ & 26  & $2.96\pm0.76$ & ~9.0 &P\\
	DEM L241 & $6.57\pm1.46$ & 0.13 &  $49.8\pm14.9$ & ~4.5 & L \\
	DEM L249 &  $5.07\pm0.93$ & 0.046 & $110\pm30.0$ & ~8.7 & L \\
	DEM L316B & $3.41\pm0.66$ &  0.17 &  $20.7\pm5.73$ &  ~7.5 &  L \\
	DEM L316A & $2.84\pm0.53$ & 0.12 & $24.0\pm6.55$ & 16.2 &L \\
	0548--70.4 & $2.04\pm0.34$ & 0.27  & $7.55\pm1.96$ & ~7.5 & P \\
	\hline
	\\
	\multicolumn{6}{l}{Galactic Sources} \\
	\hline
	Kepler	& ~131 (1) & 64.4 & 2.0 & 27.6 (8)&   D\\
	G11.2--0.3	& 1100 (2)  & 395 &  2.8 & ~6.7 (9)& D\\
	G15.9+0.2	& ~530 (2) & 95 &5.6 & 10.4 (10) &  D\\
	Kes 73	& ~990 (2) &175 & 5.6 & ~5.8 (11) & D\\
	Kes 75  	& ~270 (2) & 19.5 & 13 & ~8.0 (12) & D \\
	3C 391 	& 2000 (2) & 61.7 & 32& ~6.5 (13)&  L\\
	W44  	& ~200 (2) & 8.31 &  240 & ~5.8 (8)& L\\
	3C 396 	& ~300 (2) & 18 & 16 & ~8.1 (14) & L\\
	3C 397  	& ~810 (2) & 129 & 6.2 & ~2.9 (15) &   P\\
	W49B  	& 3000 (2) & 587 & 5.1 & ~2.8 (8) & L\\
        Cygnus Loop	& 7900 (3)& 1400 & 5.6 & ~2.7 (8) & D\\
	Cas A	& 2570 (4)  & 1870 & 1.3 &17.4 (16) & D\\
	Tycho	& ~668 (1) & 176 & 3.9 & 9.9 (8) &  D\\
	IC 443	& 8100 (3) & 1930 & 4.2 & 2.2 (8) &L \\
	Puppis A	& 9000 (5) & 3800 & 6.9 & ~7.0 (17)&  D\\
        G292.0+1.8	& ~287 (6) & 204 & 1.4 & ~7.7 (18)& D\\
	Kes 17	& ~455 (7) & 5.4 & 84 & ~8.8 (19) &L \\
	RCW 103 	& 2700 (2) & 1700 & 1.6 & ~3.5 (20) & D\\
	G349.7+0.2	& 1400 (2) & 30.3  & 42 & ~8.8 (21) & P 
	\enddata
	\tablecomments{}
	\tablenotetext{a}{Total IR flux ($F_{\rm IR}$) of the 19 LMC SNRs and the 19 Galactic
	SNRs.
	For all LMC SNRs except three, $F_{\rm IR}$ is derived using the $Spitzer$ 24 and 70
	\micron~fluxes from \citepalias{seok13} in this work.
	$F_{\rm IR}$ of 0509-67.5 and 0519-69.0 are derived from their dust mass and temperature
	measured based on their IRS spectra \citep{bwill11,borko06}.
	Uncertainties of their fluxes incorporate the dust temperature range in \citet{borko06}.
	For SN 1987A, see below.
	For the Galactic SNRs, $F_{\rm IR}$ is taken from the references given in parentheses.
	}
	\tablenotetext{b}{Soft band (0.3-2.1 keV) X-ray flux from the $Chandra$ SNR Catalog.
	Uncertainties are assumed to be 30\% en bloc (see Section \ref{subsec:obs}),
	which are taken into account for estimating the uncertainties of IRXs. 
	For two of the LMC SNRs, {\it XMM-Newton} fluxes are adopted;
	Assuming the distance of 50 kpc, we derive the X-ray flux of DEM L205 
	based on its X-ray luminosity measured by \citet{maggi12}. 
	For DEM L241, the X-ray flux of the entire SNR is derived by the summation of the X-ray
	fluxes for its head and tail structures and a point source inside the SNR \citep{bamba06}.
	For four Galactic SNRs, their X-ray fluxes are not included in the $Chandra$ Catalog,
	or the flux integrated over the limited area is only listed. 
	In these cases, we took the X-ray fluxes from the literature;
	Cygnus Loop and Puppis A: \citet{dwek87a}, IC 443: \citet{kawasaki05}, Kes 17: \citet{gelfand13}
	}
	\tablenotetext{c}{Gas temperature ($T_e$) in 10$^6$ K. For the LMC SNRs, $T_e$ is taken
	from \citepalias{seok13}. 
	For the Galactic SNRs, $T_e$ is adopted from the references given in parentheses.
	}
	\tablenotetext{d}{Spatial resemblance between IR and X-ray morphologies 
	(D: definite, P: partial, L: lack of resemblance).}
	\tablenotetext{e}{Due to the youth of SN 1987A, IR and X-ray fluxes evolve with time. 
	We adopt ranges of the measured fluxes and the corresponding IRXs from \citet{dwek10}
	between day 6067 to 7983, and the representative value of the IRX is 2.5. }
	
	\tablerefs{References for $F_{\rm IR}$ and $T_e$ of the Galactic SNRs.
	(1) \citet{gomez12}; (2) PG11 and references therein; (3) \citet{dwek87a}; (4) \citet{sibth10};
	(5) \citet{arendt10}; (6) \citet{lee09}; (7) \citet{lee11};
	(8) \citet{kawasaki05}; (9) \citet{roberts03}; (10) \citet{reynolds06}; (11) \citet{kumar14};
	(12) \citet{helfand03}; (13) \citet{chen04}; 
	(14) \citet{su11}; (15) \citet{safi05}; (16) \citet{hwang12}; (17) \citet{hwang08}; (18) \citet{park04}; 
	(19) \citet{gelfand13}; (20) \citet{gotthelf97}; 
	(21) \citet{lazendic05}
	}
\label{tab:flx}
\end{deluxetable*}

\subsection{Dust and gas cooling function}\label{subsec:theory}

The dust cooling function ($\Lambda_{\rm d}$,\,erg\,cm$^3$ s$^{-1}$)
in thermal equilibrium was computed by \citet{dwek87}, and we
adopt the simple analytic expression of the collisional heating rate from
\citet{dwek92}. 
For a single species of dust grains with a given radius ($a$ in \micron),
the cooling function is thus written as
\begin{equation}
\Lambda_{\rm d}(T)\simeq2.15\times10^{-21}\left(\frac{\mu Z_{\rm d} }{\rho a}\right)T_6^{3/2}
 h(a,T)
\label{eq:lam_d}
\end{equation}
where $\mu$ is the mean atomic weight of the gas ($\mu=1.4$),
$Z_{\rm d}$ is the DGR,
$T_6$ is the gas temperature in units of $10^6$ K,
$\rho$ is the material density of the dust grain, and
$h(a,T)$ is the heating efficiency. The heating efficiency is the fraction of
electron kinetic energy deposited to dust, which is a function of electron energy
and grain size (and species). It was extensively calculated by \citet{dwek87}, and we adopt
the approximate formula given by \citet[see their equation 6]{smith96}.
We take $Z_{\rm d}=0.0062$ from \citet[BARE-GR-S model]{zubko04}
for the Galaxy and $Z_{\rm d}=0.0017$ from \citet{meixner13} for the LMC.
We consider graphite as a common dust species for the Galaxy and the
LMC \citep[$\rho=2.24$ g cm$^{-3}$,][]{zubko04} although taking a different
dust species such as silicate does not alter our results much (see Section \ref{sec:irx_comp}).

\begin{figure}[tbp] 
\epsscale{1.2}
\plotone{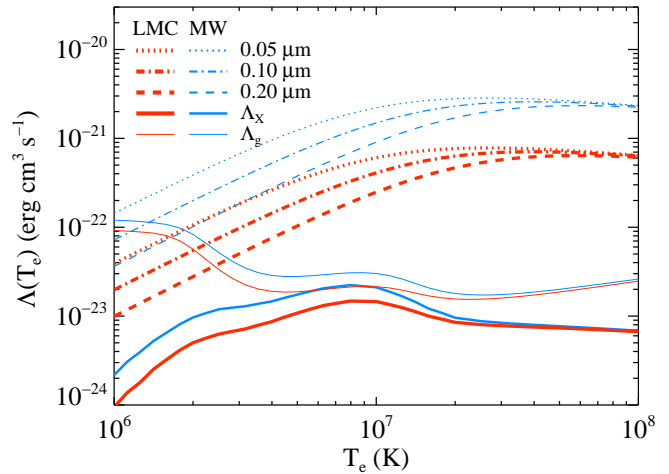}
\caption{Cooling function of a dusty plasma via gas--grain collisions for a
single-sized dust population with radii of 0.05 (dotted line), 0.10 (dot-dashed line), 
or 0.20 \micron~(dashed line) from Equation (\ref{eq:lam_d}). 
For comparison, cooling curves via atomic processes in the soft X-ray (0.3--2.1 keV) regime
($\Lambda_{\rm X}$: thick solid line), and total gas cooling curves
($\Lambda_{\rm g}$: thin solid line)
are overlaid.
Red lines represent cooling functions for the LMC while blue lines are
for the Galactic conditions. 
\label{fig:cooling}}
\end{figure}

Figure \ref{fig:cooling} shows the dust cooling functions for a single-sized dust 
populations with radii of 0.05, 0.1, and 0.2 \micron~(equation \ref{eq:lam_d}).
$\Lambda_{\rm d}$ for the Galaxy in Figure
\ref{fig:cooling} shows good agreement with $\Lambda_{\rm d}$
from more elaborate calculations by \citet{dwek08}.
In addition, gas cooling functions are shown in the figure for comparison.
For the gas cooling, we compute the cooling curve in collisional ionization equilibrium (CIE) by using
the CHIANTI code \citep[V7.0;][]{landi12}. The elemental abundances of the Galaxy and the LMC
are taken from \citet{asplund09} and \citet{russell92}, respectively. Note that we calculate not only
the total gas cooling but also the soft X-ray (0.3--2.1 keV) cooling curve,
$\Lambda_{\rm X}$, 
which we adopt for more direct comparison with the $Chandra$ X-ray fluxes.

Using the cooling function of a dusty plasma via gas--grain collisions, $\Lambda_{\rm d}(T)$, 
and of the gas via atomic processes, $\Lambda_{\rm X}(T)$, the theoretical ratio of the IR cooling
to X-ray cooling, hereafter referred to as the IRX$_{\rm th}$ ratio, is defined as
\begin{equation}
{\rm IRX_{th}~ratio}\equiv\Lambda_{\rm d}(T)/\Lambda_{\rm X}(T).
\end{equation} 
It is worthwhile to note that there are several caveats in comparing the ratio of theoretical cooling
rates to the observed IRX ratios. One would result from the fact that heavy elements are repeatedly
included when we calculate the gas and dust cooling rates.
Behind the SNR shock wave, dust grains are destroyed and some fraction of heavy elements
might be returned to the gas phase. So in SNRs, the DGR would be lower while the metal abundance
in the gas phase would be higher than those in the general ISM.
We assume the non-depleted LMC (solar) abundance for the gas cooling,
considering that the X-ray emitting gas will have an abundance close to it.
If some fraction of heavy elements are locked into dust,
the gas cooling curve will be lower than those in Figure \ref{fig:cooling}.
We will further discuss other caveats in Section \ref{sec:comp_th}.

\begin{figure*}[tbp]
\epsscale{1.}
\plotone{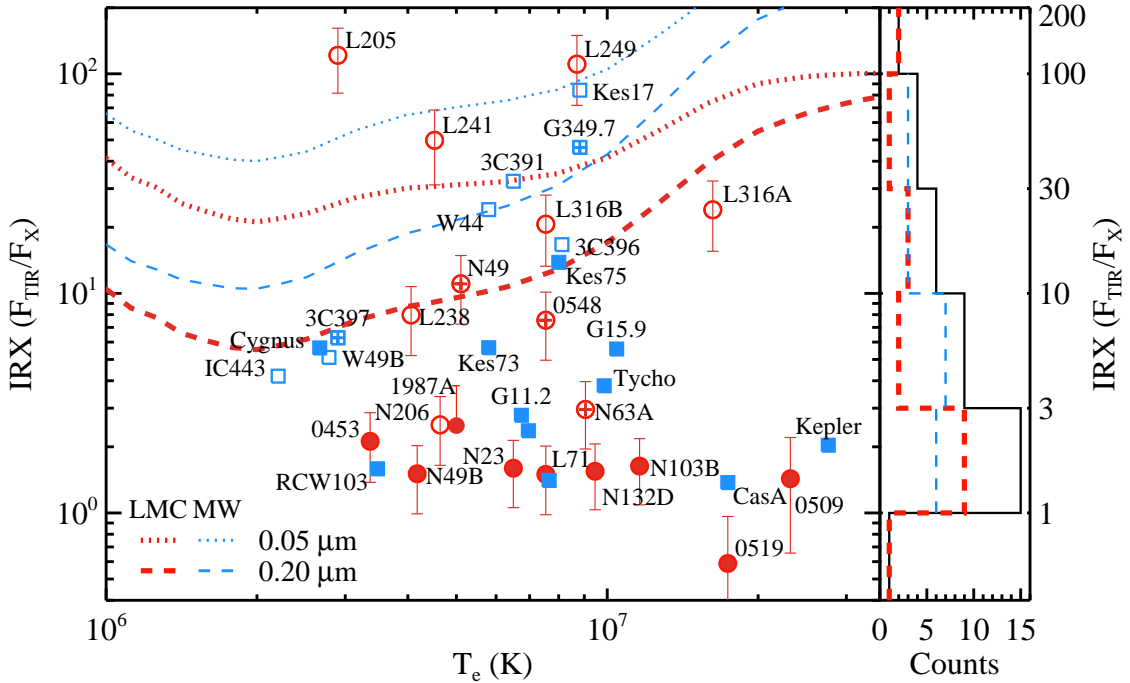}
\caption{Left: IRX ratios of LMC (circles) and Galactic (squares) SNRs as a function of plasma temperature ($T_e$).
Based on their IR and X-ray morphologies, the SNRs are classified into three groups with definite, partial,
and lack of resemblance designated by solid, open with cross, and open symbols, respectively.
Owing to lack of space, two Galactic SNRs (Puppis A and G292.0+1.8)
are not named in the figure.
Theoretical IRX ratios for single-sized dust populations 
with radii of 0.05 (dotted lines) and 0.20 \micron~(dashed lines) are overlaid for the LMC (thick red)
and Galactic (thin blue) conditions (see section \ref{subsec:theory} for the explanation).
Right: a histogram showing the IRX distributions. The IRX distributions of the LMC and the Galactic
SNRs are denoted (thick red and thin blue dashed lines, respectively) together with the total IRX
distribution (solid line).
\label{fig:irx}}
\end{figure*}

\section{Results}\label{sec:result}

Figure \ref{fig:irx} shows the IRX ratios of the 19 LMC SNRs and 19 Galactic SNRs as a function of
plasma temperature and the distributions of the IRX ratios.
The IRX resemblance is also denoted for each SNR in the figure:
definite, partial, and lack of IRX resemblance are shown respectively
by solid, open with cross, and open symbols.
Based on $\Lambda_{\rm d}$ and $\Lambda_{\rm X}$ in Figure \ref{fig:cooling},
the IRX$_{\rm th}$ ratio 
for the LMC and the Galaxy is shown in Figure \ref{fig:irx}.

First of all, the SNRs in both galaxies (except 0519--69.0) show that 
IRX ratios greater than unity, which supports
the idea that dust cooling is the primary cooling mechanism
in the dusty plasma. However, it is found that more than half of the LMC SNRs have IRX ratios just
between 1 and $\sim3$ (see also Table \ref{tab:flx}).
Moreover, IRX ratios considerably greater than unity are mostly derived from those
without a definite IRX resemblance.
This is somewhat surprising in the sense that \citet{dwek87a} report that most of the Galactic SNRs 
have IRX ratios much larger than unity. In the case of 19 Galactic SNRs shown here,
one third of them have IRX ratios below $\sim3$,
seven have ratios between 3 and 10,
and the rest of them have ratios even larger than 10.

It is noticeable that the LMC SNRs with
a lack of the IRX resemblance appear to have generally
higher IRX ratios (median ratio of those with the partial or lack of IRX resemblance: 21) than
those with definite IRX resemblance (median: 1.5). A similar trend can be found
from the Galactic SNRs, too (median ratio: 3.8 for those with definite IRX
resemblance and 24 for the rest).
These SNRs with high IRX ratios (or without definite IRX resemblance) are
mostly known to interact with molecular clouds \citep[e.g., 3C 391, W44, or IC 443,][]{jiang10} and/or 
are so-called ``mixed-morphology'' SNRs with centrally brightened X-ray emission \citep{rho98}.
For SNRs interacting with molecular clouds, the IR emission can be enhanced
because radiation may play an important role in heating dust grains
compared with collision; this eventually results in high IRX ratios.
This will be further explored for the Galactic SNRs in our future paper.

Among the SNRs with definite IRX resemblance, the IRX ratios of the LMC SNRs seem to be
systematically lower than those of the Galactic SNRs (recall the median IRX ratios of 1.5 and 3.8,
respectively). Such a difference is, in fact, consistent with the trend seen in the theoretical expectation
that the IRX$_{\rm th}$ ratio for the LMC is lower by a factor of $\sim2$--3 than that for the Galaxy
although we need to be cautious about the direct comparison between the measured IRX
and IRX$_{\rm th}$ ratios. 
The comparison between the observed IRX and IRX$_{\rm th}$ ratios is relevant
only to those
with definite IRX resemblance because the rest of them might have contributions from
radiatively heated dust. We compare only those IRX ratios with the theoretical prediction.
As shown in Figure \ref{fig:irx},
all the LMC SNRs with
definite IRX resemblance have much lower IRX ratios than IRX$_{\rm th}$
ratios by up to an order of magnitude.
Such a large deviation has been previously reported for a few Galactic
SNRs, too \citep[e.g.,][]{dwek87a}, and this could be attributed to the low DGR or dust destruction
(see Section \ref{sec:comp_th}).

\section{Discussion}\label{sec:disc}

\subsection{IRX Ratios of SNRs in the LMC and in the Galaxy}\label{sec:irx_comp}

While theoretical calculations predict that dust cooling plays a primary role
in the cooling of a dusty plasma,
the relative importance of dust cooling ($\Lambda_{\rm d}$)
to gas cooling ($\Lambda_{\rm X}$)
is expected to depend on environmental properties such as DGR, metallicity, and grain
compositions.
Our sample shows that the measured IRX ratios of the LMC SNRs are systematically lower than those
of the Galactic SNRs, and 
the theoretical IRX ratios can explain this observed trend to some extent.
The IRX$_{\rm th}$ ratio of the LMC is lower than that of the Galaxy (Figure \ref{fig:irx}),
which is due to the low DGR of the LMC. As shown in Figure \ref{fig:cooling},
while $\Lambda_{\rm X}$ at $T\ga10^6$ K is reduced by $\lesssim50\%$
due to the half-metallicity of the LMC (mainly Fe, Mg, and O),
$\Lambda_{\rm d}$ of the LMC is four times lower than that of the Galaxy
due to the 1/4-DGR of the LMC. The measured DGR of the LMC might
have some variation depending on methods or observational data
\citep[e.g.,][]{meixner10}, but nevertheless the lower IRX ratios of the LMC SNRs relative to those of the
Galactic SNRs seem to be most significantly attributed to the low DGR of the LMC.

We also examine other possibilities that may systematically influence IRX ratios.
Grain properties such as composition or size distribution may be different
between the LMC and the Galaxy though we assumed the same
properties for both. \citet{pei92} examined the extinction curves of the LMC and
the Galaxy and suggested that dust in the LMC is composed predominantly of
silicate while the abundances of silicate and carbonaceous grains are comparable in the Galaxy.
\citet{meixner10} also suggested that the standard properties for Galactic dust are not
appropriate and that amorphous carbon is required for the LMC instead of graphite.
However, the densities of different grain species do not significantly alter the dust
cooling function (Equation (\ref{eq:lam_d})), which can be a factor of two at most \citep[e.g.,
$\rho=1.81$--1.87 and 3.5 g cm$^{-3}$ for amorphous carbon and silicate, respectively;][]{zubko04}.
In addition, the details of the grain size distribution in the LMC are not well known,
but the relative abundance of very small grains to large grains in the LMC is rather similar to
the Galactic value \citep{bernard08}. This may indicate that there is no significant difference
between the LMC and the Galaxy in terms of the grain size distribution.
Therefore, the different grain composition and size distribution could be the
subsidiary factors to the low IRX ratios of the LMC SNRs.

\subsection{IRX ratio and IRX resemblance}

The SNRs without definite IRX resemblance are found to have high IRX ratios.
This may imply that there is some (or more) contribution to the IR emission originating from
radiatively heated dust or atomic/molecular emission lines rather than collisionally heated dust.
For the radiatively heated dust, there are two aspects. In the postshock gas,
radiative heating might dominate dust heating.
Using the $Spitzer$ data, \citet{ander11} modeled the dust emission of 14 Galactic SNRs and
compared the relative importance of collisional heating to radiative heating as a function of grain
size (see Figure 5 in their paper). In most cases, it is found that radiative heating is dominant
for grains with a size of 4.0--110 nm.
This is not surprising given the fact that these SNRs show signs
of interaction with molecular clouds,
which induces radiative shocks to produce the strong UV radiation
from the postshock gas \citep[e.g.,][]{DM93, koo14}.
Another aspect is that dust grains in the preshock gas can be heated by
shock radiation \citep[e.g.,][]{lee11}. If the preshock medium is dense gas such as a large molecular
cloud, then a significant fraction of the shock radiation will be absorbed and converted into IR radiation
whereas the X-ray emission from shocked gas is only a fraction of shock radiation. 
This can naturally explain the high IRX, and 
in this case we do not expect to see any spatial resemblance between IR and X-ray emission.

Additionally, ionic and/or molecular line emission could contribute to $F_{\rm IR}$.
For example, in N49, one of the SNRs that show strong emission lines in the IR spectrum,
the emission lines can contribute $\sim38\%$ of the MIPS 24
\micron~flux, and the [\ion{O}{1}] 63 \micron~line provides $11\%$ of the MIPS 70
\micron~flux \citep[and references therein]{otsuka10}.
However, since this fraction is not enough
to attribute the high IRX solely to the contribution from the line emission,
the line emission could be a relatively minor cooling source
as concluded by \citet{ander11}.

\subsection{Comparison to the theoretical IRX}\label{sec:comp_th}

Figure \ref{fig:irx} clearly shows that the LMC SNRs with a definite IRX resemblance,
have a significant deviation of the IRX ratios from the IRX$_{\rm th}$ ratios.
\citet{dwek87a} found a similar discrepancy from a few Galactic SNRs such as Kepler
or SN 1006. Both these SNRs are located far above the Galactic plane and have much
lower IRX ratios than the predicted values. 
This may suggest that the severe dearth of dust in the ambient medium around the SNRs results
in the discrepancy. 
\citet{borko06} and \citet{bwill06} calculate dust destruction in four Type Ia
and four core-collapse LMC SNRs, respectively, which overlap with our sample. 
They infer the total dust masses swept up by the blast wave before sputtering, and
these masses are generally several times lower than the typical value 
from the average DGR of the LMC. Interestingly, 0453--68.5 (and 0548--70.4),
which has a high
DGR of $9.8\times10^{-4}$ (and $7.5\times10^{-4}$),
comparable to the canonical value of
the LMC, shows a relatively high IRX ratio whereas the rest of them that have lower DGRs show
much lower IRX ratios (and consequently a more significant deviation between IRX and IRX$_{\rm th}$ ratios).
This implies that the low IRX ratios may arise from the local fluctuation of DGR
in the preshock medium of the SNR.

Another major reason for the low IRX ratios might be because dust grains are destroyed behind the
shock front and therefore the DGR decreases with time. Dust grains swept up by SNR shocks are
destroyed by thermal and non-thermal sputtering with ions. At gas temperatures above
$\sim10^6$\,K and shock speeds $\ga 500$ km s$^{-1}$, the sputtering rate is approximately constant, and the
characteristic lifetimes of carbonaceous (C) and silicate (Mg$_2$SiO$_4$)  grains with radius $a$
in \micron~are $a/|da/dt|\approx 1\times 10^6$ and $2\times 10^5 a/n_{\rm H}$ years, respectively,
where $n_{\rm H}$ is the density of H nuclei of hot plasma \citep[see Figure 2 in][]{nozawa06, dwek08}.
Therefore, for old SNRs, we expect theoretical IRX ratios lower than the curves in Figure \ref{fig:irx}.
For young SNRs, there is yet another factor. Since it takes time for the swept-up ions to reach the
CIE, the X-ray cooling rate of young SNRs is much higher than that
in the CIE model \citep[e.g.,][]{dwek87a,smith96}. Therefore, for both young and old SNRs, we expect
them below the equilibrium IRX$_{\rm th}$ ratio in Figure \ref{fig:irx}. It is anticipated
that theoretical IRX curves taking into account the dust destruction and the non-equilibrium ionization
cooling effects will provide a better understanding of the observed IRX ratios.

\subsection{Uncertainties of IRX ratios}
\label{sec:unc}

While we carry out the systematic comparison of IRX ratios in the two galaxies, 
there are some uncertainties in measuring IRX ratios that we can improve
if better knowledge of X-ray emission and more comprehensive IR analysis of SNRs
are achieved. 
The X-ray emission in an SNR may consist of different physical components such as
thermal emission from swept-up ISM or synchrotron emission.
In most cases, the $Chandra$ X-ray spectra can be explained by the
swept-up ISM with a single-temperature component, which is generally soft ($\la1$ keV).   
We have searched available literature to check the X-ray contribution from SN ejecta
or multiple components that are unlikely to be associated with the regions where dust cooling occurs.
For some SNRs (e.g., N103B, N132D, SN 1987A, and N49), however,
their X-ray spectra definitely require multiple components or a contribution from ejecta
\citep[e.g.,][]{lewis03,park03,borko07,dwek08}.
In such a case, the X-ray flux can be corrected,
as \citet{dwek08} applied a correction factor
to disentangle hard and soft components in the X-ray emission of SN 1987A,
but in general it is difficult to separate it out.
Thus, due to the contribution from SN ejecta,
IRX ratios will be lower than IRX$_{\rm th}$ ratios,
in particular for young SNRs. Indeed, as shown in Figure \ref{fig:irx}, young SNRs with
ejecta-dominated X-ray fluxes are all located in the lower part of the diagram
(e.g., \citealp[N103B:][]{lewis03}, \citealp[SNR 0509-67.5 and 0519-69.0:][]{bwill11},
\citealp[Cas A:][]{hwang12}, \citealp[Kepler:][]{reynolds06}).

While we estimate the total IR fluxes ($F_{\rm IR}$), fluxes at wavelengths
longer than 70 \micron~are not taken
into account. If there is a significant far-IR (FIR) emission
from cold dust in the SNR, we might underestimate the total IR flux. For example, \citet{graham87}
include $IRAS$ 100 \micron~fluxes for their estimation.
For the two SNRs (N49B and N63A) showing some differences between the new and previous
measurements of IRX ratios (Section \ref{subsec:obs}), their $IRAS$ fluxes at 100 \micron~are
the strongest among the fluxes in the four $IRAS$ bands. However, since both SNRs probably have
considerable FIR confusion due to nearby bright sources
\citepalias[e.g., the photoionized western lobe
embedded in N63A and a bright clump near the southern limb of N49B,
see Figures 4--5 in ][]{seok13}, their previous IRX ratios might be overestimated.

Recently, the {\it Herschel Space Observatory} has provided
FIR data with unprecedented high sensitivity and spatial resolution. 
\citet{otsuka10} detect the bright FIR emission from N49 using the $Herschel$ images and show
that a significant emission originates from the cold dust component.
In spite of its superb capabilities, however,
only two SNRs, N49 and SN 1987A, have been clearly detected by $Herschel$ so far
\citep{otsuka10,matsu11}. Note that FIR emission of SN 1987A
originates not from the swept-up ISM/CSM but from the newly formed dust \citep{remy14} whereas
most LMC SNRs in our sample are rather mature SNRs ($\ga10^3$ years).
\citet{lakicevic15} present the FIR atlas of the LMC SNRs using the $Herschel$ data and
report that SNRs are not clearly distinguishable in the FIR bands. Moreover, no
evidence for large amounts of newly formed SN-origin dust in SNRs is found, and they argue that
cold interstellar dust in SNRs is significantly destroyed due to sputtering by SN shocks. 
In this context, it is less likely that the FIR emission has a
significant influence on IRX ratios in this work
although we cannot rule out the subtle contribution.

\subsection{Implications for low-metallicity galaxies} 

The observational trend that IRX ratios of the LMC SNRs are systematically lower than those of
the Galactic SNRs implies that dust grains in extreme low-metallicity galaxies may not be a
dominant coolant of hot plasma unlike in normal-metallicity galaxies.
If the DGR is significantly reduced as the metallicity decreases \citep[e.g.,][]{lisenfeld98,hiro99},
the dust cooling would be reduced almost proportionally to the DGR. 
Meanwhile, the gas cooling is modestly affected in hot plasma ($\ga5\times10^6$ K)
in proportion to metallicity.
For example, the SMC has a metallicity even lower than
the LMC \citep[1/5-1/8 $Z_\sun$,][]{russell92}. There are more than 20 SNRs in the SMC
identified by X-ray and/or radio observations \citep[e.g., 23 SNRs listed in][]{badenes10},
and the recent {\it XMM-Newton} survey of the SMC reveals that 20 SNRs show
X-ray emission \citep{haberl12}. Surprisingly, however, only two SMC SNRs
(1E 0102.2--7219 and B0104--72.3) have been found to have the IR
counterparts \citep[e.g.,][]{stanimirovic05,koo07}. The detectability of IR emission
in the SMC SNRs is extremely low whereas about 60\% of the LMC SNRs show
associated IR emission \citepalias{seok13}. Although no systematic study of the SMC SNRs has
been carried out, this might suggest that the dust cooling in the SMC SNRs is less efficient, most
probably due to the low DGR of the SMC \citep[$Z_{\rm d}=1.1-8.3\times10^{-4}$,][]{meixner13}. 
Further IR observations toward the SMC SNRs using higher spatial resolution and higher
sensitivity such as NIRCam or MIRI onboard {\it James Webb Space Telescope} will
help in understanding the cooling mechanisms of the SMC SNRs or of hot plasma in the
low-metallicity environment in general.

\acknowledgments
This work is based on observations made with the {\it Spitzer Space Telescope},
which is operated by the Jet Propulsion Laboratory,
California Institute of Technology, under a contract with NASA.
This research was partially supported by Basic Science Research Program through the National Research
Foundation of Korea (NRF) funded by the Ministry of Science, ICT and future Planning 
(2014R1A2A2A01002811).
H. H. thanks the support from the Ministry of Science and Technology (MoST) grant
102-2119-M-001-006-MY3.

\end{document}